\documentstyle[aps,prl,epsf,floats]{revtex}
\begin{document}
\newcommand{\beq}{\begin{eqnarray}}
\newcommand{\eeq}{\end{eqnarray}}
\newcommand{\eqna}{\begin{eqnarray}}
\newcommand{\eqne}{\end{eqnarray}}
\newcommand{\dia}{\begin{displaymath}}
\newcommand{\die}{\end{displaymath}}
\newcommand{\eqnaa}{\begin{eqnarray*}}
\newcommand{\eqnae}{\end{eqnarray*}}
\def\dleft{\rlap{{\it D}}\raise 8pt\hbox{$\scriptscriptstyle\Leftarrow$}}
\def\prop{\propto}
\def\dright{\rlap{{\it D}}
\raise 8pt\hbox{$\scriptscriptstyle\Rightarrow$}}
\def\lrartop#1{#1\llap{
\raise 8pt\hbox{$\scriptscriptstyle\leftrightarrow$}}}
\def\_#1{_{\scriptscriptstyle #1}}
\def\&#1{^{\scriptscriptstyle #1}}
\def\sss{\scriptscriptstyle}
\def\dij{\delta_{\sss ij}}
\def\abs#1{\vert #1\vert}
\def\d{\delta}
\def\gij{g_{\sss ij}}
\def\Gij{g^{\sss ij}}
\def\gkm{g_{\sss km}} 
\def\Gkm{g^{\sss km}}
\def\cd#1{{}_{\sss;#1}}
\def\ud#1{{}_{\sss,#1}}
\def\udu#1{{}^{\sss,#1}}
\def\upcd#1{{}_{\sss;}{}^{\sss #1}}
\def\upud#1{{}_{\sss,}{}^{\sss #1}}
\def\ro{r\_{0}}
\def\vro{{\bf r}\_{0}}
\def\vr{{\bf r}}
\def\rar{\rightarrow}
\def\deriv#1#2{{d#1\over d#2}}
\def\oot{{1\over 2}}
\def\pdline#1#2{\partial#1/\partial#2}
\def\pd#1#2{{\partial#1\over \partial#2}}
\def\pdd#1#2#3{{\partial^2#1\over\partial#2\partial#3}}
\def\av#1{\langle#1\rangle}
\def\avlar#1{\big\langle#1\big\rangle}
\def\div{{\vec\nabla}\cdot}
\def\grad{{\vec\nabla}}
\def\curl{{\vec\nabla}\times}
\def\DD{{\cal D}}
\def\m{\mu}
\def\n{\nu}
\def\eps{\epsilon}
\def\vq{{\bf q}}
\def\dtv{d\&3v}
\def\f{\varphi}
\def\fb{\f\&{B}}
\def\eq{E_q}
\def\fl{\varphi}
\def\gf{\grad\f}
\def\b{\beta}
\def\c{\gamma}
\def\l{\lambda}
\def\L{\Lambda}
\def\om{\omega}
\def\bk{\par\noindent}
\def\a0{a_0}
\def\ao{a_0}
\def\h0{H_0}
\def\tl{T\_{\L}}
\def\va{{\bf a}}
\def\vA{{\bf A}}
\def\vD{{\bf D}}
\def\vR{{\bf R}}
\def\vF{{\bf F}}
\def\cmss{cm~s^{-2}}

\vskip 1 truein
\title{The modified dynamics as a vacuum effect} 
\author{Mordehai Milgrom}
\address{Department of condensed-matter physics,
Weizmann Institute, Rehovot Israel}



\maketitle
\begin{abstract}

To explain the appearance in MOND of a cosmological acceleration constant,
$\a0$, I suggest that MOND inertia--as embodied in the actions of free
particles and fields--is due to effects of the vacuum. The same vacuum
effects enter both MOND (through $\a0$) and cosmology (e.g. through
a cosmological constant $\L$). For example, a constant-acceleration
($a$) observer in de Sitter universe sees Unruh radiation of temperature
 $T\propto [a^2+\a0^2]^{1/2}$, with $\a0\equiv (\L/3)^{1/2}$, and
I note that $T(a)-T(0)$ depends on $a$ in the same way that MOND inertia does.

\vskip 9pt

\end{abstract}


\section{ Introduction}

The modified dynamics (MOND) has been propounded\cite{mond} as an alternative
 to the 
dark-matter doctrine, attributing the mass discrepancy evinced by galaxies
and galaxy systems to a breakdown of standard dynamics.
The pristine version of MOND states that dynamics is to be modified in the
limit of small accelerations, such as are found in galactic systems.
 (This is analogous to viewing quantum mechanics as a modification of classical
mechanics in the limit of small angular momenta.)
 More precisely, MOND asserts that 
non-relativistic dynamics involves the constant
$\a0$, with the dimensions of acceleration, so that in the formal limit
$\a0\rar 0$--i.e., when all quantities with the dimensions of acceleration are
 much larger
than $\a0$--standard dynamics obtains. In the opposite (MOND) limit of large
$\a0$ dynamics is marked by reduced inertia.
 One may roughly say that in this limit inertia at acceleration $a$ is
 $ma^2/\a0$, instead of the standard $ma$.
 (In the quantum  
analogue, mechanics involves some constant, $\hbar$, with 
the dimensions of angular momentum, such that classical mechanics prevails in
 the formal limit of small $\hbar$.)
 \par
The idea has been successfully tested against astronomical data (see e.g.
 \cite{san}\cite{mcde}\cite{sanver}) but still wants firm theoretical
underpinnings. The value of $\a0$ that emerges from such analyses 
($\approx 10^{-8}\cmss$) is of the order of $cH_0$ ($H_0$ is the Hubble 
constant)
 which is an acceleration of cosmological significance\cite{mond}. 
Even if $\a0$ is a fingerprint of
 cosmology on local dynamics, it is not necessarily a proxy for 
$a_{ex}\equiv cH_0$.
 There are other cosmological acceleration scales that one
can define\cite{com}\cite{ann}; e.g., $a_c\equiv c^2/R_c$, where $R_c$ is the
 curvature (spatial
or space-time), or $a\_{\L}\equiv c\L^{1/2}$, where $\L$ is the 
 cosmological constant. Today we have only an upper limit on $a_c$
that is of the order of $a_{ex}$, and recent evidence
(see e.g. \cite{per,efs}) points to a nonzero
cosmological constant of order $H^2_0$, for which  $\a0\sim a\_{\L}$.	
 So $\a0$ might be a proxy for any of these
cosmological parameters. 
The exact identification has important conceptual and phenomenological
consequences, e.g.  because the different acceleration scales
 depend differently on cosmic time.
 If we make such a connection with cosmology, we see that in a ``trivial'' 
universe--one that is flat, static, and has a vanishing cosmological 
constant--$\a0$ would vanish, and standard dynamics be restored.
 \par
In relativity theory, inertia and gravity are intertwined.
This might turn out to be true also in the ultimate theory for MOND. However,
in the undeveloped state in which the theory is now it is customary to think of 
MOND as either a modification of gravity (e.g. \cite{bm}) or as of one of
 inertia (e.g. \cite{ann}).
 The difference between the two options
is not only one of principle. One
 can expect major differences in predictions regarding the dynamics of
 astrophysical systems\cite{ann}.
\par
Here, I shall concentrate on the modified-inertia interpretation of MOND.
The line I would like to follow is that inertia is a derived property of matter
 that results from the interaction of matter with some
 agent--the vacuum, in the present case. Cosmology either affects or is
 affected by the vacuum. Inertia is thus
 influenced by, or shares common influences with, cosmology; and, it is through
 this that $\a0$ enters local dynamics and cosmology.
In other words, either cosmology has a causative effect on inertia because it 
affects the state of the vacuum, which, in turn, affects inertia, or,
 cosmology and inertia are both affected by the vacuum dynamics, which
then enters cosmology say as a cosmological constant, $\L$, and MOND through 
$\a0\approx c\L^{1/2}$.
\par
The same connection could come about in MOND as modified gravity;
 at present I deem modified inertia a more promising avenue.
\par
Inertia was introduced into physics as an attribute that endows bodies
with energy and momentum according to their motion, and these can be changed
only by subjecting the body to a force. Attempts to generate inertia--in
 the spirit of Mach's principle--have traditionally concentrated on inertia of
 bodies (see e.g. \cite{bar}). More generally, however, inertia is the 
attribution
of energy and momentum to all dynamical degrees of freedom, whether we describe
them as bodies (particles) or fields. This attribution is conventionally
encapsuled in the action of the free particles and field, from
which the energy-momentum tensor is gotten. To derive inertia
it might behoove us then to derive the free actions of matter fields.
Alternatively, we might start from the equation of motion of fields, and try to
 derive its free part as some sort of force exerted by the vacuum.
 Apart from occasional comments I shall limit myself in what follows
to the discussion of the archetypal case of inertia of bodies.
\par
In \S II I discuss the modified-inertia
formulations of MOND. In \S III I discuss various points that might be
 pertinent to inertia as a vacuum effect, and in \S IV
I add some comments that are specific to MOND inertia.


\section{MOND as modified inertia}

 \par
Assuming that MOND is
underpinned by an action--which is not at all obvious--a possible intermediate
 stage in formulating MOND as modified inertia is to look for an effective
kinetic action that governs the motion of a body. 
 If $\vr(t)$ is the 
trajectory of the body, we want the kinetic action to be of the form

\beq A_mS[\vr(t),a_i], \label{gtarata} \eeq

such that $A_m$ depends only on the body, and $S$ depends only on the trajectory
$\vr(t)$ and on the parameters $a_i$ measuring the departure from standard 
physics, and presumably connected with cosmology.
 The coefficient $A_m$ can then be identified as 
the rest mass of the body (gravitational and inertial) as the body's 
contribution to the energy-momentum tensor will be $A_m t^{\m\n}$ with 
$t^{\m\n}$ depending only on the trajectory. In the limit of ``trivial''
cosmology, we would want $S$ to go to the standard expression $\int d\tau$.
The interaction action remains intact by MOND according to this picture. 
\par
In general, all aspects of cosmology might enter $S$.
 However, almost all the astrophysical systems studied in 
light of MOND to date (galaxies of all sorts, galaxy groups, and clusters) are
 characterized by  very nonrelativistic motions, by
 sizes much smaller than the the cosmological scale, and by dynamical times
 smaller than the Hubble time. 
For such systems the approximate statement of MOND would
be that cosmology is connected to local
 dynamics only through the one parameter $\a0$,
which is of the order of the accelerations found in these systems.
The approximate effective action is thus written as
\beq A_mS[\vr(t),\a0]. \label{gtaputa} \eeq
This, as I said, need not be the action for bodies still
partaking in the Hubble flow, or that are relativistic; there, cosmology might
enter in a more elaborate manner. In the limit $\a0\rar 0$, $S$ should go
to the standard expression
 $\lim_{T\rar \infty}T^{-1}\int_{-T}^{T}v^2/2~dt$.
In the opposite limit, MOND phenomenology dictates that
  $S$ scale as $\a0^{-1}$. In particular, an important tenet of MOND is that 
inertia vanishes as $\a0\rar\infty$ (see below for an explanation of this
in our Machian scheme).
\par
Actions of the form (\ref{gtaputa}) have been studied in detail in \cite{ann}.
 It was shown, for instance, that Galilei invariance, combined with the 
two limits in $\a0$,
 requires that $S$ be a nonlocal (in time) functional of the trajectory.
This should not be viewed as an imposition. On the contrary,
 nonlocality appears
 naturally in effective action as ours must be. Also, nonlocality might be a
 blessing as nonlocal
theories need
not suffer from the maladies that are endemic to higher-order theories
(\cite{ann} and references therein). Nonlocality also opens the way to
 solving, in modified-inertia versions,
 the center-of-mass motion of composite bodies (the problem is solved in 
the modified-gravity version of ref.\cite{bm}):
 MOND inertia is nonlinear; how then do the equations of
 motion for constituents combine to give the correct equation of motion
 for the center-of-mass motion? This question comes up at the
 microscopic level of inertia-from-vacuum models, and also at the macroscopic
 level\cite{mond}. For
 instance, atoms inside stars have high accelerations, yet for MOND to
 work we want these stars, which move with low accelerations
in galaxies, to follow MOND inertial behavior.
The problem can probably not be solved in local theories (i.e.
when the inertia force depends only on a finite number of time derivatives of
the trajectory). In nonlocal theories this is not the case. Consider, for
 example, a heuristic model in which the  
equation of motion
for a particle moving in a potential $\f$ is of the form
\beq \om^2\vR(\om)\mu[\abs{\vA(\om)}/\a0]=\vD(\om), \label{lodutaba} \eeq
where we work in Fourier space at frequency $\om$, $\vR(\om)$ is the transform
of $\vr(t)$, $\vD(\om)$ is the transform of $-\gf[\vr(t)]$, and 
\beq \vA(\om)\equiv \int_{0}^{\om}\hat\om^2\vR(\hat\om)~d\hat\om
 \label{puraka}\eeq
is a measure of the accumulated acceleration up to frequency $\om$.
 (This equation of motion is not derivable from an action.)
Conventional dynamics corresponds to the choice $\mu\equiv 1$. It is 
straightforward to see that, indeed, in this theory the center of mass
 (CM) motion 
of a body is hardly affected by internal motions inside it. If the 
accelerations characterizing internal motion are higher than
 those for the CM motion
their frequencies must also be higher (because $a\sim \omega^2 r$, and $r$ is
 smaller); these internal components do not then enter $\vA$ for the CM
motion. The CM motion does enter internal dynamics strongly in what is 
known in MOND as the external-field effect\cite{mond}. One can demonstrate all
this by solving exactly equations (\ref{lodutaba})(\ref{puraka})
 for a potential that is harmonic in the 
$x-y$ plane, and linear in the $z$ direction: $\f=\Omega^2(x^2+y^2)/2+gz$. 
One finds that the $z$ motion is of constant acceleration, which corresponds
 to
frequencies near $\omega=0$, and is totally oblivious to the $x-y$ motion.
The $x-y$ motion is still harmonic but with a frequency that depends on the 
amplitude and on the $z$ acceleration (and goes to $\Omega$ in the
 high-acceleration limit) exemplifying the external-field effect.
\par
There are different MOND action of the form (\ref{gtaputa}) that one could
 write, but no one, in particular, that deserves
to be pinpointed as the ``appropriate'' action. They were all shown\cite{ann}
 to lead to an exact relation that governs
 circular motion in axisymmetric potentials (applicable to rotation curves):
\beq a\mu(a/\a0)=g_N,  \label{miolata} \eeq
where $a=v^2/r$ ($v$ and $r$ are the orbital velocity and radius) $g_N$ is
 the Newtonian
acceleration, and $\mu(x)$ is the MOND interpolating function, which here
follows from the exact form of the action as restricted to circular orbits. 
It is this relation that has always been used in analyses of galaxy rotation 
curves (e.g. in \cite{san}\cite{mcde}\cite{sanver}).
\par
How might one derive such an effective action from more fundamental concepts,
and, in particular, how does cosmology come to be connected with
  local dynamics
in the MOND limit through $\a0$?
It was suggested in \cite{ann} that this could come about if inertia is
 due--in the spirit of Mach's principle--to the effect of some agent, such as
 the vacuum, on non-inertial bodies. Since the vacuum is itself modified by 
the non-trivial state of the Universe, its inertia effect might also be 
imprinted with this information. Alternatively, it might be that it is not
cosmology that affects local dynamics, but that the same vacuum 
attribute enters both cosmology, as a cosmological constant for instance,
 and MOND as $\a0\approx c\L^{1/2}$.
\par
 In the next two sections I consider this surmise in more detail.
I can offer no specific mechanism that begets MOND inertia from vacuum effects,
but there is a number of pertinent ideas that might be of help in looking for
 such an underlying theory.


\section{Inertia as a vacuum effect}

 As stated above, one way to generate inertia from effects of the vacuum is
to derive the
contribution of free fields to the action. In conventional dynamics these are
 $ m\int d\tau$ for a
free particle, $\int F_{\m\n}F^{\m\n}$  for the electromagnetic field, 
 $\int(\partial_{\n}f\partial^{\n}\f)^2+m\f^2$ for a massive-scalar field, etc.
 MOND would contend that these are the 
approximate effective actions only in the limit of trivial cosmology,
 but, in general, effects of cosmology enter the action. In particular,
 in the nonrelativistic limit cosmology enters through $\a0$.
\par
If the free action is to be derived 
from effects of the vacuum fields, then we must start with vacuum fields
that themselves have no free action; we need to start with 
inter-field interactions only, and produce from these all the free actions.
\par
It is well known that interactions can induce, or modify, inertial actions.
For example, the effective mass of ``free'' electrons and holes in a 
semiconductor are greatly modified. 
More pertinent to our concern with the vacuum, mass renormalization 
in field theory	is, of course, a vacuum effect. The Higgs mechanism induces
an effective mass term from the interaction with the putative Higgs field.
It is also known that the interaction of the electromagnetic field with
charged vacuum fields induces a contribution to the free action of the 
electromagnetic field--the Heisenberg-Euler effective action
(see e.g. \cite{zel}, \cite{adler}, and \cite{izu} p. 195).
It is not clear whether, and exactly how,
 the mechanisms listed above fit the bill as concerns MOND.
However, those that are known to affect inertia must be reckoned with in
any complete analysis. 
\par
  Sakharov\cite {sakh} (see also \cite{adler} for a review)
 proposed a scheme to derive the ``free'' action of gravity from effects of 
the vacuum: Curvature of space-time modifies the dynamical behavior of vacuum 
fields, hence producing an
associated energy or action for the metric field. To lowest order (in the 
Planck length over the curvature radius) this gives the Einstein-Hilbert
action $\int g^{1/2}R$. Sakharov's 
arguments make use of the fact that the vacuum fields have inertia (since
they are assumed to carry the usual energy-momentum). So, derived inertia 
comes prior to induced gravity a-la Sakharov.
The two concepts are not, otherwise, related by the equivalence principle--as
incorrectly stated in \cite{rh}. The Einstein-Hilbert action is not determined
 by the weak equivalence principle; the latter enters through the matter
 action alone (in dictating, e.g. that free particles move on geodesics,
 etc.).
\par
 One can also bypass the action, start from the equation of motion, and try
to derive the free part as a force exerted by the vacuum on non-inertial
bodies. For instance,
the Casimir effect may modify the inertia of macroscopic, as well as 
microscopic, bodies. In the least, the Casimir  energy of a body should 
contribute to its inertial mass. Indeed, it was shown in \cite{jr}
that two parallel mirrors accelerated together on an hyperbolic (linear,
constant-acceleration) trajectory experience a force proportional to the 
acceleration, with a coefficient (induced mass) that corresponds to the Casimir energy of the double mirror. A Casimir energy stems from the energy-momentum
 of the vacuum fields, and so presupposes inertia of the vacuum fields.
 In a different attempt,
it has been suggested in \cite{rh} that a body in acceleration is subject to 
a flux of radiation arising from the vacuum, and that if this flux is partially
absorbed by the body, the momentum transferred might supply the inertia
force on the body.
 Even if correct, this picture also tacitly presupposes that the vacuum fields
 themselves have inertia, because it builds on their carrying momentum, which
 they lose at impact (for instance, Maxwell dynamics is assumed for the vacuum
 electromagnetic field, which already implies inertia).  
 Such mechanisms, which presuppose inertia, cannot be the 
 primary origin of inertia.  
Yet another attempt\cite{hrp} views each body as composed of charged 
oscillators that can oscillate only in the plane perpendicular to the
acceleration--and here the alignment of the inertia force with the acceleration
is put in by hand. The inertia force is then said to be a Lorentz-like force
produced by the zero-point fluctuations on the constituent oscillators.
\par
A precondition for a dynamical, inertia-producing effect of the vacuum is that
an observer be able to perceive its non-inertiality through
vacuum effects.  
It is indeed well known that
for such an observer the vacuum transforms itself into a palpable radiation
field. For an observer on a collinear trajectory of constant-acceleration, $a$,
(hyperbolic motion)
this is the Unruh radiation: a thermal bath of temperature $T=a\hbar/2\pi kc$
 \cite{unruh}\cite{birdav}.
(From here on I shall work in units where $\hbar=1$, $c=1$, $k=1$, so
$T=a/2\pi$.)
 For a more general motion, the incarnation of the vacuum is
non-thermal, and hardly anything seems to be known
 about the Unruh-like radiation.
 Some simple instances
of stationary motion have been studied, e.g., in \cite{ger}\cite{audretsch}.  In particular, circular, highly relativistic motions have been discussed in 
\cite{audretsch}-\cite{leinaas}. In this last case, as for hyperbolic motions,
 $a=\c^2 v^2/r\approx \c^2/r$ determine the spectrum of
 the avatar of the vacuum  ($\c$ is the Lorentz factor). This 
 is quasi-thermal with effective temperature $T=\eta a/2\pi$,
  where $\eta$ is of order unity and depends somewhat on the frequency.
Circular motion in general is characterized by two parameters, which can be
taken as $a$ and $r$. This leads to a rather more complicated Unruh effect,
which has not been described yet.
For stationary trajectories all points are equivalent, and so the Unruh-like
radiation can be described in terms of ``local'' properties of the trajectory.	
 Clearly, for a general motion the effect becomes a
nonlocal functional of the whole trajectory.
\par
Even at this basic level of the observer's need to read its motion in the 
vacuum 
 it is not clear that he can read in it all that it needs. For instance	can an observer in hyperbolic motion tell its direction, which would be
necessary to define an inertia force? The Unruh flux itself is said
to be isotropic \cite{ger}, although there may be other attributes that the 
observer can pick up (e.g. fluctuations).
Perhaps we need to take finite-size effects into account. For example, by 
comparing the Unruh radiation that its different parts see, the body could
 divine the direction of its acceleration.
\par
 It is thus doubtful whether
 the Unruh-like radiation itself is implicated in directly generating inertia.
 There is also the problem of response time: In 
standard dynamics inertia is simply proportional to the acceleration, but this
 may change on time scales that are shorter than the
 typical period of the Unruh radiation.
It is claimed in \cite{rh} that their mechanism does give instantaneous 
adjustment because their putative flux of vacuum quanta appears to depend only
on the instantaneous acceleration. This, however, cannot be correct if the 
acceleration changes on time scales that are not much longer than the typical
 period of the fields.
 For nonrelativistic, circular
 motion with velocity $v$ and orbital radius $r$ the
typical wavelength of the Unruh radiation that corresponds to the acceleration 
 is $\l=a^{-1}=r/v^2\gg r$, and
  the typical frequency of the radiation,
$\omega=v\Omega\ll\Omega$ ($\Omega$ is the angular frequency). So it is hard 
to see how
 the Unruh radiation itself can produce inertia that responds instantaneously 
to the the value of the acceleration, which we know must be the case at least
in the non-MONDian regime. This could happen if in the non-MOND regime
vacuum effects are dominated by the short-wavelength limit. If this is 
dictated, say, by a cutoff near the Planck scale, which is much smaller than
 any scale of relevance that we have probed so far, then instantaneous (local)
inertia will follow. This is the case in Sakharov's induced
 gravity, in the standard derivation of the cosmological constant from vacuum
effects, in the derivation of the Heisenberg-Euler effective free action for
electromagnetism, mentioned above, etc.
In contradistinction, in the MOND regime there is no experimental indication
 that inertia is instantaneous. In fact, as we saw, theoretical arguments lead
 us to expect that inertia is nonlocal. 
\par
 It is clear from the undirected account given so far that I can offer no 
specific mechanism for inertia from vacuum. One vague notion that might be
 invoked is the Le Chatelier principle: Non-inertial motion is accompanied 
by a nontrivial manifestation of the vacuum. Inertia might then be an attempt
 of the vacuum to restore equilibrium and minimize the thermal effects of
 non-inertiality. One would like to find an action functional of the trajectory
of the form (\ref{gtarata}) 
that embodies such an effect. The actual interaction with the vacuum will enter
the mass factor $A_m$, while the ``kinematic'' factor that depends on the
trajectory alone will come from the transformation of the vacuum to the
 non-inertial frame


\section{MOND as a vacuum effect}
\par
If the idea underlying MOND is basically correct, it will add a new dimension to
the Machian concept of inertia. One will
now have to explain the appearance of $\a0$ from more fundamental concepts, and
the connection with cosmology.
Beside the appearance of a cosmological parameter, MOND introduces two major 
changes: inertia becomes non-linear and, most probably, nonlocal.
\par
Can we spot an inkling of MOND inertia in the Unruh effect? 
When the acceleration of a constant-$a$ observer becomes smaller than $\a0$,
the typical frequency of its Unruh radiation becomes smaller than the expansion
rate of the Universe, the Unruh wavelength becomes larger than
the Hubble distance, etc. \cite{com}\cite{ann}. We expect then some break in the response
of the vacuum when we cross the $\a0$ barrier. It is interesting, as a 
preliminary indication, to understand the nature of the Unruh radiation
seen by a non-inertial observer in a nontrivial universe.
 It is well known that even inertial observers
in a nontrivial universe observe, generically, some sort of radiation arising
from the distortion of the vacuum.
 The simplest and best studied case is that of a de Sitter universe where all
inertial observers see a thermal spectrum with a temperature 
$\tl=(\L/3)^{1/3}/2\pi$ \cite{gibhawk}, where $\L$ is the cosmological
 constant characterizing the de Sitter cosmology. 
It was shown recently \cite{npt}\cite{des}
 that an observer on a constant-a (hyperbolic) trajectory, in a de Sitter
 universe also sees thermal radiation, but with a temperature
\beq T(a)={1\over 2\pi}(a^2+\L/3)^{1/2}.  \label{pultada} \eeq
de Sitter space-time can be embedded in a five-dimensional Minkowski space
time. In the latter, the above observer moves in exact hyperbolic motion with 
the absolute value of the 5-acceleration 
\beq a_5=(a^2+\L/3)^{1/2}.  \label{kupotara} \eeq 
The radiation that observer sees can thus be viewed as the Unruh radiation in
 the 5-D embedding space. It can be shown that relation (\ref{kupotara})
 between the instantaneous,
absolute value of the 4-acceleration in de Sitter space, and that in the 
embedding Minkowski space holds for an arbitrary trajectory.
\par
Harking back to the Le Chatelier principle,
if we envisage inertia as a force that drives a non-inertial body back to
equilibrium as regards the vacuum radiation (here, drive $T$ back to $\tl$)
 then $T-\tl$ is a relevant quantity.
We can write
\beq  2\pi(T-\tl)\equiv2\pi\Delta T= a\hat\mu(a/\hat\a0), \label{musata} \eeq
 with
\beq \hat\m(x)=[1+(2x)^{-2}]^{1/2}-(2x)^{-1},  \label{mura} \eeq
and $\hat\a0=2(\L/3)^{1/2}$.
The quantity $\Delta T$ behaves in just the manner required from MOND 
inertia\cite{mond}, with $\a0=\hat\a0$ identified naturally with a cosmic
acceleration parameter.
[Note that $\hat\mu(x\ll 1)\approx x$,
 $\hat\mu(x\gg 1)\approx 1-(2x)^{-1}$.]
The significance of this suggestive finding is anything but obvious.
First, it is not really clear why $\Delta T$ should be a measure of inertia
( similar quantities such as $T^2-\tl^2$ do not give the correct
MOND behavior). Second, it is difficult to see how to generalize the 
argument to arbitrary motions. It would be particularly interesting to 
generalize it to nonrelativistic, circular motions,
relevant to rotation-curve analysis of disk galaxies--the most clear-cut
test of MOND to date. For circular orbits, all nonlocal inertia
 theories lead to an effective inertia force of the 
form $a\mu(a/\a0)$--as in eq.(\ref{miolata})--where $\mu(x)$ need not take the
 same form as for the 
liner-motion case, but must still satisfy  $\mu(x\ll 1)\approx x$,
 and $\mu(x\gg 1)\approx 1$.
\par
Note here that the quantity $a\pd{T}{a}$, which measures, e.g., the 
temperature change under small dilatations of the orbit, also gives 
 a MOND-like expression
\beq  a\pd{T}{a}= a\mu(a/\a0), \label{kulata} \eeq
\beq \m(x)=x/(1+x^2)^{1/2},  \label{muta} \eeq
and $\a0=(\L/3)^{1/2}$, although the significance of this is, again, not clear. 
\par
In de Sitter space-time the expansion rate, the space-time
curvature, and the cosmological constant are all the same.
These parameters differ from each other in a general Friedmanian
universe (the first two are not even constant) and so the above lesson
learnt for the de Sitter case does not tell us which of the cosmological
acceleration parameters is to be identified with $\a0$ in the real Universe.
\par 
The qualitative, phenomenological tenet of MOND whereby inertia vanishes
 in the limit $\a0\rar \infty$ can be understood in the present picture
 as follows. The limit
 corresponds to $\L\rar\infty$, or 
$\h0\rar \infty$, etc.; so, the Gibbons-Hawking-like radiation
due to cosmology swamps the
thermal effects due to non-inertial motion such that the 
difference between inertial
and non-inertial observers is erased in this limit.
\par
As already noted above, the cosmological connection need not enter as a 
causative effect of cosmology on local dynamics--mediated by the vacuum).
 It may be that vacuum effects enter both cosmology--e.g. through a 
cosmological
 constant--and local dynamics through $\a0$, thus establishing the ``cosmic
coincidence '' seen in MOND.

\begin{acknowledgements}
I thank Jacob Bekenstein for many helpful discussions
\end{acknowledgements}


\begin{thebibliography}{999}



\bibitem{mond}M. Milgrom,  Astrophys. J. {\bf 270} 365 (1983)
\bibitem{san}R.H. Sanders, Astrophys. J. {\bf 473} 117 (1996)
\bibitem{mcde}S.S. McGaugh and W.J.G. de Blok, Astrophys. J. 
{\bf 499} 66 (1998)
\bibitem{sanver}R.H. Sanders and M.A.W Verheijen, Astrophys. J. 
{\bf 503} 97 (1998)
\bibitem{com}M. Milgrom, Comments on Astrophysics {\bf 13(4)}, 215 (1989)
\bibitem{ann}M. Milgrom, Ann. Phys. {\bf 229}, 384 (1994)
\bibitem{per}S. Perlmuter et al., preprint astro-ph/9812133 (1998) 
\bibitem{efs} G. Efstathiou, S. L. Bridle A. N. Lasenby, M. P. Hobson,
and  R. S. Ellis, preprint  astro-ph/9812226 (1998) 
\bibitem{bm}J. Bekenstein and M. Milgrom, Astrophys. J. {\bf 286} 7 (1984)
\bibitem{bar}J.B. Barbour and H. Pfister (eds.), {\it Mach's Principle}, 
Birkhauser (Boston) (1995)
\bibitem{zel} Ya. B. Zeldovich, Pis'ma Zh. Eksp. Teoret. Fiz. {\bf 6}, 922
(English translation in JETP Lett. {\bf 6}, 345 1967)
\bibitem{adler} S.L. Adler, Rev. Mod. Phys. {\bf 54}, 729 (1982)
\bibitem{izu} C. Itzykson and J.B. Zuber, {\it Quantum Field Theory},
McGraw-Hill (1980)
\bibitem{sakh} A.D. Sakharov, Sov. Phys. Doklady, {\bf 12}, 1040 (1968) 
\bibitem{rh}A. Rueda and B. Haisch, Physics Letters A {\bf 240}, 115 (1998)
\bibitem{jr}M.T. Jaekel and S. Reynaud, J. de Physique {\bf 3}, 1093 (1993)
\bibitem{hrp}B. Haisch, A. Rueda, and H.E. Puthoff, Phys. Rev. A
{\bf 49}, 678 (1994)
\bibitem{unruh}W.G. Unruh, Phys. Rev. D {\bf 14}, 870 (1975)
\bibitem{birdav} N.D. Birell and P.C.W. Davies, {\it Quantum Fields in
Curved Space}, Cambridge University Press (Cambridge), 1982
\bibitem{ger} U.H. Gerlach, Phys. Rev. D {\bf 27}, 2310 (1983)
\bibitem{audretsch}J. Audretsch, R. M\"uller, and M. Holzmann
Class. Quantum Grav. 12 2927 (1995)
\bibitem{bell}J.S. Bell and J.M. Leinaas, Nuc. Phys. {\bf B284}, 488
(1987)
\bibitem{lpp}O. Levin, Y. Peleg, and A. Peres, J. Phys. A: Math Gen.
{\bf 26}, 3001 (1993)
\bibitem{untwo}W.G. Unruh, preprint hep-th/9804158 (1998)
\bibitem{leinaas} J.M. Leinaas, preprint hep-th/9804179 (1998)
\bibitem{gibhawk}G.W. Gibbons and S.W. Hawking, Phys. Rev. D {\bf 15},
2738 (1977)
\bibitem{npt}H. Narnhofer, I. Peter, and W. Thirring, Int. J. Mod.
 Phys. B {\bf 10} 1507 (1996)
\bibitem{des} S. Deser and O. Levin, Class. Quant. Grav. {\bf 14}, L163 (1997)
\end{thebibliography}
\end{document}